\documentstyle[12pt]{article}
\newcommand{\asl}{\!\not\!\! A}
\newcommand{\beq}{\begin{equation}}
\newcommand{\beqarr}{\begin{eqnarray}}
\newcommand{\CS}{Chern-Simons}
\newcommand{\D}{{\cal D}}
\newcommand{\dsl}
  {\kern.06em\hbox{\raise.15ex\hbox{$/$}\kern-.56em\hbox{$\partial$}}}
\newcommand{\eeq}{\end{equation}}
\newcommand{\eeqarr}{\end{eqnarray}}
\newcommand{\ZZ}{{\rm \kern 0.275em Z \kern -0.92em Z}\;}
\begin{document}
\title     {On the Quantization of the Abelian
          \CS\ Coefficient at Finite Temperature}
%
\author{%
N. Brali\'c\thanks{nbralic@lascar.puc.cl}\\
{\normalsize\it
  Facultad de F\'{\i}sica, Pontificia Universidad Cat\'olica de Chile}\\
{\normalsize\it
  Casilla 306, Santiago 22, Chile}\\
\rule{0cm}{1.cm}
C.D. Fosco\thanks{fosco@ictp.trieste.it}\\
{\normalsize\it
  International Centre for Theoretical Physics}\\
{\normalsize\it
  P.O. Box 586, 34100 Trieste, Italy}\\
\rule{0cm}{1.cm}
F.A. Schaposnik\thanks{Investigador CICBA}\\
{\normalsize\it
  Departamento de F\'{\i}sica, Universidad Nacional de La Plata}\\
{\normalsize\it
  C.C.~67, (1900) La Plata, Argentina}%
}
\date{}
\maketitle
\begin{abstract}
We show that when the Abelian \CS\ theory coupled to matter
fields is quantized in a vacuum with non vanishing magnetic
flux (or electric charge), the requirement of gauge invariance
at finite temperature leads to the quantization of the \CS\
coefficient and its quantum corrections, in a manner similar
to the non-Abelian case.
\end{abstract}
\newpage
\CS\ gauge theories have been the subject of great attention 
during the past decade~\cite{ss,djt}.  In the non-Abelian case,
the interest in pure \CS\ theories stems from their topological
character, whereas the Abelian \CS\ gauge field is responsible
for inducing generalized statistics on matter fields, a phenomenon
intrinsic to three-dimensions.

A distinctive feature of \CS\ theories is that the requirement
of gauge invariance leads to non perturbative effects prior to
quantization.  In the non-Abelian case, gauge transformations
fall into topological classes labeled by $\pi_3(G) = \ZZ$,
corresponding to the mapping of $S_3$ (compactified
three-dimensional space) to the group manifold $G$.  Under
gauge transformations with non-vanishing winding, the \CS\
action $S_{CS}$ is not invariant, and requiring the invariance
of $\exp(iS_{CS})$ leads to the quantization of the non-Abelian
\CS\ coefficient~\cite{djt}.  In the Abelian case there is no
such topological structure, so the Abelian \CS\ coefficient
remains arbitrary, and free to induce any desired generalized
statistic on the matter fields.

When the theory is formulated at finite temperature, the time
direction is effectively compactified into a circle.  Since the
Abelian group manifold is also a circle, it is tempting to ask
whether the Abelian \CS\ coefficient remains arbitrary at finite
temperature.  In this letter we analyze that question, and show
that under certain general conditions the Abelian \CS\ coefficient
is indeed quantized at {\it any\/} finite temperature.  However,
this quantization turns out to be of a different origin and
character as compared to the non-Abelian case.  It requires the
coupling with matter fields, and results from the interplay of
the winding of gauge transformations around the compactified
time axis, and the winding of the (pure gauge) potential at
spatial infinity.  The latter determines the total flux of the
magnetic field, and in the presence of matter fields specifies
the vacuum of the theory.  Our result states that when the
theory is defined in a vacuum with non vanishing magnetic flux
(or electric charge), the requirement of gauge invariance leads
at finite temperature to the quantization of the \CS\ coefficient
and its quantum corrections.  Indeed, the situation is similar
to that encountered at zero temperature when the theory is
formulated in an appropriately compactified spacetime
manifold~\cite{polyNP,hosotani}

Our starting point is the action for massive Dirac fermions coupled
to an Abelian \CS\ field at finite temperature
\beq
  S = S_F + i \frac{\theta e^2}{4\pi} S_{CS} \;,
\label{S_class}
\eeq
where $S_{CS}$ is the \CS\ action
\beq
S_{CS} = \int_0^\beta d\tau \int d^2 x \,
         \epsilon_{\mu\nu\lambda} A_\mu \partial_\nu A_\lambda 
\label{S_cs}
\eeq
and $S_F$ is the fermion action
\beq
  S_F = \int d^3 x \, \bar\psi (\dsl + i e \asl + m) \psi \;.
\label{S_f}
\eeq

Finite temperature calculations are done as usual, compactifying
the (Euclidean) time variable $\tau$ into the range 
$0 \le \tau \le \beta = 1/T$ (in our units, $\hbar=c=k=1$).
Then, the partition function is defined as
\beq
{\cal Z} = {\cal N}(\beta) \int \D \bar\psi \D \psi \D A_\mu \>
           \exp\left( -S \right) \;,
\eeq
where the functional integral must be computed using periodic
(antiperiodic) boundary conditions in time for bosons (fermions).
It is important to stress that the integration over gauge fields
ranges over all periodic configurations compatible with the
boundary conditions at spatial infinity to be discussed below.
Using the standard Fadeev-Popov procedure, one gets
\beq
{\cal Z} = {\cal N}(\beta)
           \int \D \bar\psi \D \psi \D A_\mu \D \chi \>
           \delta[F[A]] \Delta_{FP}[A] 
           \exp\left( -S[\bar\psi,\psi,A^\chi] \right) \;.
\label{Z}
\eeq
Here $\Delta_{FP}$ is the Fadeev-Popov determinant associated
with the gauge-fixing condition $F[A]=0$, and $\D\chi$ is the
integration over the group of gauge transformations.  The
action of these gauge transformations on the fields is given by
\beqarr
  A_\mu(\tau,x) &\to& A^\chi_\mu(\tau,x)
                   = A_\mu(\tau,x) + \partial_\mu \chi(\tau,x)
  \nonumber\\
  \psi(\tau,x) &\to& \exp[-ie\chi(\tau,x)] \psi (\tau,x)
  \nonumber\\
  \bar\psi(\tau,x) &\to& \exp [+ie\chi(\tau,x)] \bar\psi(\tau,x) \;.
\label{GaugeTransf}
\eeqarr
In order to preserve the temporal boundary conditions for both
bosonic and fermionic fields, the function $\chi$ must satisfy
\beqarr
  \partial_\mu\chi(\beta,x)  -  \partial_\mu \chi(0,x) &=& 0 \\
  \exp[-ie\chi(\beta,x)] -  \exp [-ie\chi(0,x)] & = & 0 \;.
\label{exp-chi}
\eeqarr

Let us write the partition function as
\beq
{\cal Z} = {\cal N}(\beta) \int \D A_\mu
           \delta[F[A]] \Delta_{FP}[A] 
           \exp\left( -S_{eff}[A] \right) \;,
\label{Z-eff}
\eeq
where we have defined the effective action $S_{eff}[A]$ by
\beq
\exp(-S_{eff}[A])
  = \int \D \bar\psi \D\psi \D\chi
    \exp\left( -S[\bar\psi,\psi,A^\chi] \right) \;,
\eeq
or, after integration over the fermion fields,
\beq
\exp(-S_{eff}[A]) =
  \int \D \chi
  \exp \left(i\frac{\theta e^2}{4\pi} S_{CS}[A^\chi] \right)
  \times \det (\dsl + ie \asl^\chi + m)  \;.
\label{S_eff}
\eeq
The fermion determinant has been profusely analyzed, both at
zero and at finite temperature~\cite{det}-\cite{Det}.  Various
perturbative treatments coincide in showing that the determinant
has a parity-violating part, to be discussed below, and a
parity-conserving piece which contains the Maxwell action.
The presence of the latter is important here, since it fixes
the boundary conditions for the gauge fields at spatial infinity.
In order for a field configuration to have finite energy, the
field strength must vanish at infinity, and therefore the gauge
potentials must tend to pure gauges.

Evidently, the effective action $S_{eff}[A]$ defined  by
eq.(\ref{S_eff}) is gauge invariant: $S_{eff}[A^\eta] = S_{eff}[A]$
for any $\eta$ satisfying the appropriate boundary
conditions. Moreover, if the \CS\ action and the fermion determinant
were gauge invariant, the integration over the gauge group would
factor out, and this effective action would reduce to the usual one
obtained after integrating over the fermion fields.  Here, however,
we shall see that, even in this Abelian theory, there is a gauge
dependence due to `large' gauge transformations.  Thus, $S_{eff}$
will receive a non trivial contribution from the integration over
gauge transformations, which precisely restores gauge invariance,
even under large gauge transformations.

The set ${\cal G}$ of all gauge transformations is partitioned
into topological classes ${\cal G}_n$, classified by an integer
$n$ denoting how many times they wind around the time coordinate,
according to
\beq
  \chi^{(n)}(\beta ,x) - \chi^{(n)}(0,x) = \frac{2\pi}{e} n \;.
\label{chi_n}
\eeq
A representative of ${\cal G}_n$ may be taken as a gauge
transformation defined by a function $\chi^{(n)}$ of the form
\beq
\chi^{(n)}(\tau ,x) =
  \frac{2\pi}{e} \frac{n\tau}{\beta} + \chi^{(0)}(\tau ,x) \;,
\eeq
where $\chi^{(0)}(\beta,x)=\chi^{(0)}(0,x)$.  Such elements
obviously satisfy~(\ref{exp-chi}) and~(\ref{chi_n}).  These
transformations have been considered in the case of non-compact
QED$_3$ in Ref.~\cite{gss}, where  their relevance to the study
of confinement was discussed.

The integration over gauge transformations in eq.~(\ref{S_eff}) can
be written as a sum over integrations within each topological sector,
to get
\beq
\exp(-S_{eff}[A]) =
  \sum_n \int \D \chi^{(n)}
  \exp\left( i\frac{\theta e^2}{4\pi} S_{CS}[A^{\chi^{(n)}}] \right)
  \times \det (\dsl + ie \asl^{\chi^{(n)}} + m) \;.
\eeq
To proceed further we must determine the way in which the integrand
in this expression transforms under `large' ($n \ne 0$) gauge
transformations.  Concerning the \CS\ action, from eq.~(\ref{S_cs})
one easily verifies that
\beq
S_{CS}[A^{\chi^{(n)}}] =
  S_{CS} [A] + \delta_{\chi^{(n)}} S_{CS} [A] \;,
\label{gauge-tr-S_cs}
\eeq
where
\beqarr
\delta_{\chi^{(n)}} S_{CS} [A]
  &=&
    \int_M d^3 x \, \epsilon_{\mu\nu\lambda}
    \partial_\mu \chi^{(n)} \partial_\nu A_\lambda
  =
    \int_M d^3 x \, \partial_\mu (\epsilon_{\mu\nu\lambda}
	\chi^{(n)} \partial_\nu A_\lambda ) \nonumber\\
  &=&
    \int_0^\beta d\tau \int d^2 x \,
	\left[
      \partial_\tau ( \epsilon_{jk} \chi^{(n)} \partial_j A_k )
      + \partial_j ( \epsilon_{jk} \chi^{(n)} F_{k0})
	\right] \;.
\label{delta1}
\eeqarr
The second integral in this expression vanishes due to the boundary
condition at infinity discussed earlier for the field strength.
For the first term, we have
\beqarr
\int_0^\beta d\tau \int d^2 x \,
  \partial_\tau ( \epsilon_{jk} \chi^{(n)} \partial_j A_k )
  &=&
    \int d^2 x [ \chi^{(n)}(\beta,x) \,
    \epsilon_{jk} \partial_j A_k (\beta,x) \nonumber\\
  & &\quad
    - \chi^{(n)} (0,x) \, \epsilon_{jk} \partial_j A_k(0,x) ]
	\nonumber\\
  &=& \int d^2 x [\chi^{(n)}(\beta,x) - \chi^{(n)}(0,x)]  
    \epsilon_{jk} \partial_j A_k(0,x) \nonumber\\
  &=& \frac{2 \pi n}{e} \int d^2 x \,
    \epsilon_{jk} \partial_j A_k(0,x) \nonumber\\
&=& \frac{2 \pi n}{e} \Phi (0) \;,
\label{delta2}
\eeqarr
where the boundary condition (\ref{chi_n}) and the periodicity
of the gauge fields have been used.  Here, $\Phi(\tau)$ denotes
the magnetic flux through the spatial plane at `time' $\tau$:
\beq
  \Phi(\tau) = \int d^2 x \, \epsilon_{jk} \partial_j A_k(\tau,x) \;.
\label{flux}
\eeq
which can be written in terms of the gauge potential at spatial
infinity as
\beq
  \Phi(\tau) = \oint dx^i A_i(\tau,x) \;.
\label{stokes}
\eeq
As discussed earlier, the gauge potential in the integrand of this
expression must be a pure gauge.  Thus, taking into account the
transformation law of the fermion fields in eq.~(\ref{GaugeTransf}),
we see that the flux $\Phi$ must be of the form
\beq
  \Phi(\tau) = \frac{2\pi}{e} q  \;,
\label{Phi_q}
\eeq
where, in general, $q$ is an integer-valued function of $\tau$.
However, field configurations with different fluxes at different
`times' $\tau$, will be separated by barriers of infinite energy.
Since the Abelian theory has no instantons~\cite{inst} that may
connect those configurations, we must conclude that the value of
$\Phi$ in eq.~(\ref{Phi_q}) is $\tau$-independent, and consequently,
$q$ is a fixed integer, which labels the vacuum of the theory and
determines the boundary conditions for the functional integral
defining the partition function.  In other words, each value of
$q$ defines a different theory, in a way analogous to the different
vacua of theories with spontaneous symmetry breaking.  The partition
function corresponding to each of these theories will be denoted
by ${\cal Z}_q$ and, from eq.~(\ref{Z-eff}), is given by
\beq
{\cal Z}_q = {\cal N}(\beta) \int \D A^{(q)}_\mu \>
             \delta[F[A^{(q)}]] \, \Delta_{FP}[A^{(q)}] 
             \exp\left( -S_{eff}[A^{(q)}] \right) \;,
\label{Z_q}
\eeq
where $\D A^{(q)}_\mu$ denotes the functional integral measure
subject to the boundary condition specified in eq.~(\ref{Phi_q}).

The condition on the total flux in the functional integral~(\ref{Z_q})
can be converted into a condition on the total charge
$Q = e \int d^2x\,{\bar \psi}\gamma_0 \psi$, due to the constraint
\beq
Q (\tau) = - \frac{\theta e^2}{2 \pi} \Phi (\tau) \;,
\eeq
which follows from~(\ref{S_class}).  Thus, condition~(\ref{Phi_q})
implies
\beq
Q = - \theta e q 
\eeq
i.e., ${\cal Z}_q$ in~(\ref{Z_q}) can be equivalently regarded
as a  partition function in a fixed-charge ensemble.  Such
partition functions, where conserved internal quantum numbers
are given prescribed values, have been considered previously in
the literature~\cite{redl,kapu}.  The application of this idea to
the present case amounts to Fourier transforming the partition
function with an imaginary chemical potential (see~\cite{kapu}
for details).  This may turn out to be a practical procedure
to enforce the condition~(\ref{Phi_q}) in the functional integral
for ${\cal Z}_q$.

Coming back to the gauge dependence of the \CS\ action, we must
note that when the theory is quantized with $q \ne 0$, the simple
discussion of $\delta_{\chi^{(n)}} S_{CS}$ in
eqs.~(\ref{delta1}-\ref{delta2}) is not quite correct.  Indeed, in
that case the gauge potential $A^{(q)}$ cannot be defined globally
without introducing singularities or, alternatively, one must define
$A^{(q)}$ resorting to several coordinate patches.  As shown in
refs.~\cite{alvarez,polyNP}, that leads to an extra factor of two in
eq.~(\ref{delta2}), which together with eqs.~(\ref{gauge-tr-S_cs})
and~(\ref{Phi_q}) lead to
\beq
  \delta_{\chi^{(n)}} S_{CS}[A^{(q)}] = \frac{(2\pi)^2}{e^2} 2nq  \;.
\label{delta-S_cs}
\eeq

We now discuss the way in which the fermion determinant changes
under large gauge transformations.  An exact expression for
the determinant is unknown, but perturbative analysis~\cite{Det}
suggest that its parity-violating part is proportional to the
\CS\ action, with a temperature dependent coefficient:
\beq
\det (\dsl + ie \asl + m) \vert_{PV} =
    \exp\left(i\frac{e^2}{4\pi} F(T) S_{CS}[A] \right) \;.
\label{Det-Pert}
\eeq
where the subindex $PV$ denotes the parity-violating contribution.
Concerning the known parity-conserving terms, they are gauge
invariant even under large gauge transformations, and will
play no role in the analysis that follows.  Then, assuming
that eq.~(\ref{Det-Pert}) holds, we get
\beq
\exp(-S_{eff}[A^{(q)}]) =
  \sum_n \int \D\chi^{(n)}
  \exp
  \left(
    i\frac{e^2}{4\pi} (\theta+F(T)) S_{CS}[A^{(q)\chi^{(n)}}]
  \right) \;.
\eeq
Then, using eqs.~(\ref{gauge-tr-S_cs}) and~(\ref{delta-S_cs}),
we have
\beqarr
\exp(-S_{eff}[A^{(q)}]) &=&
  \exp\left( i\frac{e^2}{4\pi} (\theta+F(T)) S_{CS}[A^{(q)}] \right)
  \nonumber \\
  & & \times
  \sum_n \exp\left( i 2\pi (\theta + F(T)) qn \right) \;,
\eeqarr
where we have omitted an infinite normalization constant arising
from the volume of integration over the gauge group.  Thus, we have
\beqarr
\lefteqn{\exp(-S_{eff}[A^{(q)}]) =} \nonumber \\
  & &
  \exp\left( i\frac{e^2}{4\pi} (\theta+F(T)) S_{CS}[A^{(q)}] \right)
  \sum_k \delta\left( (\theta + F(T)) q  - k \right)  \;.
\eeqarr
Hence, the partition function vanishes, unless
\beq
 (\theta + F(T)) q = p  \;,
\label{quant}
\end{equation}
where $p$ is an integer-valued function of $T$.

For the theory in a $q \ne 0$ sector, this relation yields a
novel quantization rule.  Since it must hold even if we start
with $\theta = 0$, this result states that $F(T)$, the
coefficient of the \CS\ term induced by the fermionic quantum
fluctuations at finite temperature, must be a rational-valued
function of the temperature.  This, in turn, implies that in
the theory at finite temperature, $\theta$ itself must be a
rational number.  (Actually, the denominators in these rationals
is $q$, which is fixed).  Notice that our argument does not
exclude a possible temperature-dependence of $F(T)$:  it only
states that at most it can be a rational-valued function of
$T$.  

Perturbative analysis leading to a temperature dependence
for the fermion determinant of the form~(\ref{Det-Pert}), make
at some point in the calculation a $1/m$ expansion, and
produce a function $F(T)$ of the form
\beq
  F(T) = \frac{1}{2} \tanh \left( \frac{m\beta}{2} \right) \;.
\eeq
Those perturbative calculations have been performed only in
what we have denoted here as the $q=0$ sector, where we obtain
no conditions for $\theta$ or $F(T)$.  Yet, it is interesting
to note that in the limit $m \to \infty$, where the $1/m$
expansion could be expected to be exact, this expression tends
to a step function, up to exponentially small corrections, thus
satisfying the quantization condition~(\ref{quant}).

It is interesting at this point to compare the situation in
the Abelian and the non-Abelian models.  For the latter, it
is well known that the \CS\ coefficient has to be quantized
already at zero temperature if $\exp(iS_{CS})$ is to be invariat
under large gauge transformations~\cite{djt}.  In contrast, no
quantization condition arises at $T = 0$ for the Abelian theory
(unless the spatial dimensions are compactified~\cite{polyNP,hosotani}.)
For the non-Abelian theory, it was argued in ref.~\cite{cfrs}
that the \CS\ coefficient must remain an integer at any finite
temperature.  Thus, any temperature-dependent renormalization
of the non-Abelian \CS\ coefficient must reduce to an integer
shift.  Here we have seen that an analogous result holds also
in the Abelian theory:  in any $q \ne 0$ sector, a
temperature-dependent renormalization of the \CS\ coefficient
must satisfy condition~(\ref{quant}), so it cannot be a smooth
function of the temperature.  This will be of importance for
the applications of the theory to anyonic superconductivity,
where the cancellation of the bare \CS\ coefficient against its
quantum corrections plays a crucial role~\cite{AnionicSC}.  A
quantization condition like eq.~(\ref{quant}) guarantees that
that cancellation will hold at least up to some finite non-zero
temperature, thus surviving to low temperature effects.  Also,
one may expect that as in the non-Abelian case, the quantization
of the \CS\ coefficient will provide a non-perturbative basis
for its non-renormalization to higher orders~\cite{ColemanHill}.
This, in turn, is important for the applications of the theory to
the Fractional Quantum Hall effect~\cite{FQH}.  

As opposed to the non-Abelian case, in the Abelian theory the
presence of fermions (or scalars, for that matter) is essential
for the argument leading to the quantization of the \CS\
coefficient at finite temperature.  Indeed, it is the presence
of matter fields that forces conditions~(\ref{exp-chi})
and~(\ref{chi_n}) on the allowed gauge transformations, and
determine the behavior of the gauge fields at spatial infinity,
leading to the quantization of the magnetic flux by the integer
$q$ in eq.~(\ref{Phi_q}).  As discussed above, different values
of $q$ label theories built over different classical vacua, there
being no tunneling between the different $q$-sectors.  We have
shown that in each $q$-theory the Abelian \CS\ coefficient is
quantized at finite temperature in multiples of $1/q$, except
for $q=0$ where our analysis imposes no conditions on $\theta$
or $F(T)$.

In topological terms, what we have shown is that the quantization
of the abelian \CS\ coefficient at finite temperature, results
from the interplay of two different $\pi_1(U(1))=\ZZ$.  First,
there is the $\pi_1(U(1))$ whose elements $q$ label the different
classical vacua, and determine the boundary conditions at spatial
infinity defining the partiton function ${\cal Z}_q$ in
eq.~(\ref{Z_q}).  Then, to enforce the invariance of ${\cal Z}_q$
under `large' gauge transformations, we summed over the integers
in the $\pi_1(U(1))$ which classify the gauge transformations
allowed at finite temperature, as specified in eq.~(\ref{chi_n}).
As a result, for $T>0$, the coefficient of the \CS\ term is
quantized according to eq.(\ref{quant}) in each $q \ne 0$ sector.
It is interesting to note that these conclusions hold for
{\it any\/} $T>0$, and therefore will hold also in the $T \to 0^+$
limit.

\medskip
\underline{Acknowledgements}.  This work was supported in part by
FONDECYT, under Grant No. 1950794, by CONICET, under Grant PID
3049/92, the ICTP and by Fundaci\'on Andes and Fundaci\'on Antorchas.
FAS thanks the Pontificia Universidad Cat\'olica de Chile for its
kind hospitality.

%
\end{document}